# Minority Carrier Diffusion Lengths for High Purity Liquid Phase Epitaxial GaAs

D. Alexiev, D. A. Prokopovich, L. Mo

Australian Nuclear Science and Technology Organisation, Menai, N.S.W. 2232, Australia.

## Abstract
The diffusion length of minority carriers $L_{p,n}$ is an important characterisation parameter in semiconductor materials and is of particular interest when constructing devices such as solar cells ( Hovel 1975 ), double hetero junction lasers ( Casey and Panish 1978 ) and bipolar transistors. Their efficiency depends primarily on the ability of minority carriers to diffuse through neutral material to a p-n junction or Schottky barrier where they recombine with majority carriers. For this reason diffusion lengths have been measured in a variety of semiconductor materials.
The GaAs material was grown by liquid phase epitaxy (LPE) at the Australian Nuclear Science and Technology Organisation. The diffusion lengths measured for high purity p-type and n-type LPE-GaAs samples were observed to be longer than any previously reported. Measurements of minority carrier diffusion lengths for p-type and n-type GaAs were carried out using an electron beam induced current (EBIC) technique.

## 1. Introduction
The liquid phase epitaxial (LPE) GaAs examined here was grown for possible construction of room temperature semiconductor detectors of X-rays and soft γ-rays. For these devices a Schottky barrier is constructed on a 100-300 μm thick GaAs epitaxy; the depletion region formed by the Schottky barrier is the detector's sensitive region since majority carriers, created as electron-hole pairs during the impact ionisation initiated by incoming photons, are collected with high efficiencies in this region of the detector.
The reason for measuring $L_{p,n}$ in these majority carrier devices was to confirm the quality of the material grown. Low carrier concentrations are required for the formation of a depletion region of adequate depth for the detection of penetrating radiation, but freedom from carrier recombination centres is a further requirement since this improves energy resolutions and charge collection properties. Radiation detectors made so far from the LPE material have shown good energy resolution, qualitatively indicating high purity material. Also, deep level transient spectroscopy (DLTS) measurements have shown no detectable deep level recombination centres, again indicating high purity material. The $L_{p,n}$ measurements can establish semiconductor purity because the presence of recombination centres within an epitaxial layer will result in reduced values of $L_{p,n}$. These centres are associated with a variety of defects and impurities in the material, including Ga vacancies (Ettenberg et al. 1976) and residual impurities such as oxygen and transition metals as reported by Jastrzebski et al. (1979). However, of particular concern are large concentrations of non-radiative recombination centres in the melt-grown substrate (Sekela et al. 1975) which are used for the growth of the epitaxial layers. Jastrzebski et al. (1979) reported that under particular conditions, influenced by growth parameters and growth mode, out-diffusion may occur from the substrate into the epitaxial layer. For this reason $L_{p,n}$ is of immediate interest as a means of assessing the quality of an epitaxial layer.

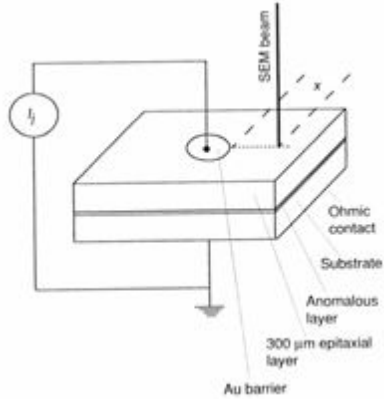

**Fig. 1.** Schematic of GaAs sample showing the electron beam path and the current measurement path.

## Measurement of $L_{p,n}$

The measurement of $L_{p,n}$ using electron beam induced currents (EBIC) is based on work by Higuchi and Tamura (1965) and Wittry and Kyser (1965) who found $L_{p,n}$ values for p-n junctions. A recent review of EBIC methodology using scanning electron microscopes (SEM) was given by Holt (1989), though the method used here was first suggested by Thornton (1968) and Hackett et al (1972). Briefly, an electron beam incident normal to a Schottky diode is slowly scanned across the surface of the sample. No bias is applied to the diode and the short circuit current induced by the electron beam is measured as a function of the distance x between the beam and the edge of the surface diode, as shown in Fig. 1.

When the electron beam penetrates a specific distance into the epitaxial layer, a point source of excess electrons and holes is established; these then diffuse through the zero-field region of the epitaxy. If the material is n-type, holes reaching the zero-bias depletion region under the Schottky contact are rapidly collected; if the epitaxy is p-type then the opposite applies, with electrons being collected.

It is more usual for p-n junctions or Schottky barriers to have the electron beam scanned parallel to the junction, as described by Leitch et al. (1981); the resulting current $I_j$ varies with distance x from the Au barrier as

$$I_j \propto \exp(-x/L_{p,n}). \qquad (1)$$

Loannou and Dimitriadis (1982) have examined the alternative case of scanning the beam normal to the plane of the Schottky barrier. They found that in the presence of an infinite surface recombination velocity S, $I_j$ obeys the relation

$$I_j \propto x^{-3/2} \exp(-x/L_{p,n}). \qquad (2)$$

Kuiken and van Opdorp (1985) extended this treatment to include the effect of finite values of S. However, both those reports assumed a zero-bias barrier depletion width much smaller than the depth of the point source of minority carriers generated in the semiconductor by the electron beam. This approximation did not hold for the low carrier concentration samples studied here since they had zero-bias depletion regions of the order of the depth of the point source. The large depletion regions provided a charge collection plane parallel to the electron beam in a geometry which approaches that of a p-n junction held parallel to an electron beam. It is therefore reasonable to expect a situation where Ij is described by equation (1) at least to a first-order approximation. Thus if a straight line results from a plot of $\ln |I_j|$ versus x, then the slope will equal $-1/L_{p,n}$. Corresponding values of $\tau_{p,n}$ will be given by

$$\tau_{p,n} = L^2_{p,n}/D_{p,n}, \qquad (3)$$

where $\tau_{p,n}$ is the carrier lifetime, $D_{p,n}$ is the hole or electron diffusion coefficient $[=(kT/q)\mu_{p,n}]$, and $\mu_{p,n}$ is the hole or electron mobility.

## 3. Experimental

The SEM used for these experiments was a JEOL JXA840. The beam distance from the Au barrier was calibrated by simply photographing the sample and relating its magnification bar (in μm) to the beam position on the screen of the SEM system. Particular attention had to be paid to shielding the internal signal lead from stray current pickup due to secondary electron scattering. The diode (surface barrier) circuit was earthed at one point only; the current measuring instrument, a Keithley 614 electrometer, was left floating (see Fig. 1).

Sample preparation followed standard procedures: a chemically cleaned section of LPE-GaAs is etched, masked and a 2 mm diameter Au area on the epitaxy and an Al surface on the substrate are evaporated to form the surface barrier and ohmic contact respectively.

## 4. Results

A number of p- and n-type LPE-GaAs samples of various net carrier concentrations were examined at different beam voltages. EBIC versus distance plots, measured at 300 K, are shown in Figs 2 and 3.

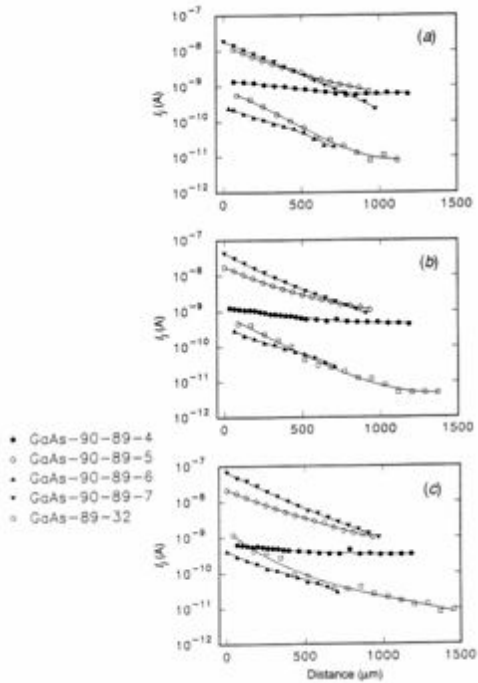

**Fig. 2.** Measured EBIC currents for n-type GaAs samples using different electron beam voltages: (a) 15 kV, (b) 25 kV and (c) 35 kV

For all but one of the samples (GaAs-89-5.3-2) the initial decrease in the current $I_j$ closely followed equation (1), but was then followed by a slower decrease at larger distances, apparently representing a second diffusion length $L_2$ notably longer than the first. Such changes in slope for GaAs samples have also been noted by Ryan and Eberhardt (1972) and Wittry and Kyser (1965).

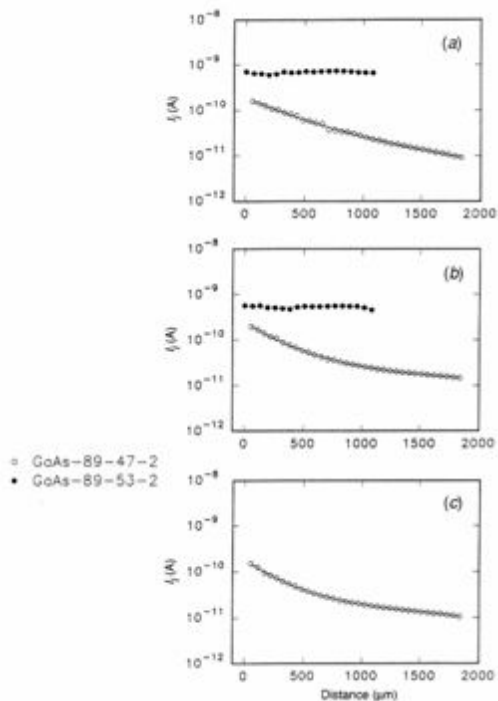

**Fig. 3.** Measured EBIC currents for p-type GaAs samples using different electron beam voltages: (a) 15 kV, (b) 25 kV and (c) 35 kV

The processes determining currents at larger distances from the barrier (or p-n diffused junction) were tentatively explained by Wittry and Kyser (1965) as being the result of majority carrier production by infrared radiation absorbed near the Schottky barrier. This phenomenon was originally predicted by Dumke (1957). Holt and Chase (1973) confirmed the explanation of Wittry and Kyser (1965) and have further found that the value of $L_2$ is dependent on x; so that at x = 1000 μm, for instance, the observed value of $L_{p,n}$ would be ~1000 μm.

Table 1. Measured and theoretical values of minority carrier diffusion length $L_p$ and lifetime $\tau$ for n-type GaAs

| Specimen | $N$ (cm$^{-3}$) | Theoretical values | | Measured $L_p$ ($\mu$m) | | |
|---|---|---|---|---|---|---|
| | | $L_{dr}$ ($\mu$m) | $\tau_{dr}$ ($\mu$s) | 15 kV | 25 kV | 35 kV |
| GaAs-90-89-4 | $1\times10^{13}$ | 770 | 600 | 830 | 780 | 920 |
| GaAs-90-89-5 | $2\times10^{14}$ | 174 | 29 | 270 | 260 | 260 |
| GaAs-90-89-6 | $1\times10^{14}$ | 250 | 62 | 270 | 280 | 280 |
| GaAs-90-89-7 | $1\cdot4\times10^{14}$ | 208 | 42 | 230 | 210 | 200 |
| Ryan & Eberhardt | $6\times10^{13}$ | 320 | 100 | 100 | 200 | 200 |
| Alferov et al. | $5\times10^{15}$ | 34 | 1·1 | | 11·1 | |
| Wittry & Kyser | $5\cdot1\times10^{16}$ | 10 | 0·1 | | 4 | |
| Tansley | $2\times10^{18}$ | 1·8 | 0·003 | | 1·6 | |
| Hwang | $2\times10^{16}$ | 18 | 0·3 | | 1·5 | |

Table 2. Measured and theoretical values of minority carrier diffusion length $L_n$ and lifetime $\tau$ for p-type GaAs

| Specimen | $N$ (cm$^{-3}$) | Theoretical values | | Measured $L_n$ ($\mu$m) | | |
|---|---|---|---|---|---|---|
| | | $L_{dr}$ ($\mu$m) | $\tau_{dr}$ ($\mu$s) | 15 kV | 25 kV | 35 kV |
| GaAs-89-53-2 | $5\times10^{14}$ | 510 | 12 | — | — | — |
| GaAs-89-47-2 | $5\times10^{14}$ | 510 | 12 | 490 | 340 | 300 |
| Wu & Wittry | $5\times10^{17}$ | 16 | 0·012 | | 1·2 | |
| Leitch et al. | $4\times10^{18}$ | 5·7 | 0·0015 | | 3·1 | |
| | $1\times10^{17}$ | 36 | 0·059 | | 9·6 | |
| | $2\times10^{17}$ | 25 | 0·029 | | 3–14 | |

Since $I_j$ was affected by extraneous phenomena at larger a;, all slope calculations were based on $\ln |I_j|$ versus x data at values of x found near the barrier (up to < 500μm). It is believed that this methodology was conservative and would actually tend to underestimate $L_{p,n}$. Values of $L_{p,n}$ with related net carrier concentrations N, found from C-V measurements, are shown in Table 1 for n-type GaAs and in Table 2 for p-type GaAs. Both tables include, for comparison, published values of $L_{p,n}$.

## 5. Limiting Values of Lp,n

The upper limit of $\tau_{p,n}$ can be set by the high probability of direct radiative recombination $B_{dr}$ [equation (10) of Hall (1959)]:

$$B_{dr} = 0.58 \times 10^{-12} \bar{n} \left(\frac{m}{m_n + m_p}\right)^{\frac{3}{2}} \left(1 + \frac{m}{m_n} + \frac{m}{m_p}\right) \left(\frac{300}{T}\right)^{\frac{3}{2}} W_G^2$$

where $\bar{n}$ is the refractive index, $m_n$, $m_p$ are the density of states effective masses of electrons and holes, and $W_G$ is the band gap.
For GaAs at 300 K we have
$\quad = 3.6$
$\quad m_n = 0\text{-}068$
$\quad m_p = 0.5$
$\quad W_G = 1.45$ eV
Therefore
$\quad B_{dr} = 1.6 \times 10^{-10}$ cm$^3$s$^{-1}$.
$\quad$ lifetime $\tau_{dr} = 1/B_{dr}\cdot n$
where n is the net carrier concentration.

Consequently it will be noted from this relationship the lifetime ($\tau_{dr}$) increases as the net impurity level of the material decreases. Tables 1 and 2 include derived limiting values for $L_{p,n}$ (and $\tau_{p,n}$) as a means of comparing experimental results.

## 6. Discussion

Results obtained from EBIC measurements of LPE-GaAs show that the minority carrier lifetimes measured at 300 K are near the limiting theoretical values. To achieve this the material examined has to be of high purity, with low levels of non-radiative recombination centres. When comparing the material used for the Ryan and Eberhardt (1972) $L_p$ measurements to material grown during this study, it appears to be of similar high purity. A further inquiry into the origin of their LPE-GaAs (Eberhardt et al. 1971; Hicks and Manley 1969) reveals that the epitaxial layer came from a Spectrosil silica crucible / Spectrosil furnace reaction tube growth system, similar to the arrangement used in our study. Further evidence for the high purity of the LPE material grown in our study comes from DLTS measurements of the epitaxial layers (Alexiev et al. www.arxiv.org ID: cond-mat/0408653). Those measurements examined the material over the temperature range from 380 to 11 K, and showed no deep level traps at a sensitivity of $N_T$ >. $10^{11}$ cm$^{-3}$, confirming the high purity of the material. The use of the LPE material for X-ray and $\gamma$-ray detection also qualitatively indicates that the material is of high purity, with resolutions of 2-3 keV full width half maximum being achieved for 60keV $\gamma$-rays at room temperature.

The flat response of $\ln |I_j|$ versus x noted in sample GaAs-89-53-2 in Fig. 3 may have been the result of recombination at the site of a crystal defect or barrier breakdown due to the proximity of defects. Examination of the sample used showed surface abnormalities, including lineages near the barrier, which could contribute to this effect. At one point, approximately 50 to 100 μm from the barrier, a sharp drop was noted in the measured value of $I_j$ to about 10% of the values shown in Fig. 3. Although the line of measurement to the barrier of sample GaAs-89-53-2 avoided this point, it was obvious that a large crystalline structure defect was present near the barrier. The high current otherwise shown by this sample may have been the result of diffusion of the minority carriers to the weak p-n junction formed between the epitaxy and the substrate (which was of course kept at a constant distance from the point source throughout the measurement).

The EBIC technique used for the measurement of the $L_{p,n}$ values employed geometrise which have not been well treated by theoretical studies. For our samples the electron beam was scanned perpendicular to the plane of the Schottky barrier (see Fig. 1), a situation treated by Loannou and Dimitriadis (1982) and others for the specific case of zero depletion width. However, because of the large depletion widths of these samples, the carrier collection area was of the order of the electron beam penetration. A rigorous theoretical analysis of this situation is beyond the objective of this paper, however, the problem was addressed by comparing the measured data with the limiting cases provided by two existing models. The first limiting case is that of an electron beam scanned perpendicular to the plane of a Schottky barrier with zero depletion width; the second case is that of an electron beam scanned parallel to an infinite charge collection area—as occurs when scanning an electron beam parallel to a p-n junction. It was found that the second limiting model best fits our data, with the first model overestimating the values of

$L_{p,n}$. To the accuracy expected here this treatment was adequate, though of course more rigorous models could improve the data analysis.

The effect of surface recombination was thought to have had minimal influence on the values of $L_{p,n}$ found here. This is evidenced by the consistency of the measurements when different electron beam energies were used. For lower electron beam energies the point source of minority carriers generated in the semiconductor is closer to the surface. Therefore, if surface recombination had been a dominant feature of these devices, it would have caused a notable lowering of the measured values of Lp,n compared with the values found using a higher electron beam energy.

We note that the values of $L_{p,n}$ measured are the highest so far reported for LPE-GaAs, indicating the high quality of the material produced.

## Acknowledgements

The authors would like to acknowledge K.S.A. Butcher and T.L. Tansley for their valuable contribution in discussions related to this work. This work was done in part for the PhD. Thesis of D. Alexiev published in 1990.